\documentclass[prl,amsfonts,amssymb,twocolumn,showpacs]{revtex4}
\usepackage{mathrsfs}
\usepackage{graphicx}
\usepackage{hyperref}
\usepackage{CJK}

\def\be{\begin{equation}}
\def\ee{\end{equation}}
\def\ba{\begin{eqnarray}}
\def\ea{\end{eqnarray}}

\begin{document}

\title{Pseudo Parity-Time Symmetry in Optical Systems}

\author{Xiaobing Luo$^{1,2}$}

\author{Jiahao Huang$^{1}$}

\author{Honghua Zhong$^{1}$}

\author{Xizhou Qin$^{1}$}

\author{Qiongtao Xie$^{1,3}$}

\author{Yuri S. Kivshar$^{4}$}

\author{Chaohong Lee$^{1,4}$}

\altaffiliation{Corresponding author. Electronic address: chleecn@gmail.com}

\affiliation{$^{1}$State Key Laboratory of Optoelectronic Materials and Technologies, School of Physics and Engineering, Sun Yat-Sen University, Guangzhou 510275, China}

\affiliation{$^{2}$Department of Physics, Jinggangshan University, Ji'an 343009, China}

\affiliation{$^{3}$Department of Physics and Key Laboratory of Low-Dimensional Quantum Structure and Quantum Control of Ministry of Education, Hunan Normal University, Changsha 410081, China}

\affiliation{$^{4}$Nonlinear Physics Centre, Research School of Physics and Engineering, Australian National University, Canberra ACT 0200, Australia}

\date{\today}

\begin{abstract}
We introduce a novel concept of the {\em pseudo} parity-time ($\mathcal{PT}$) symmetry in periodically modulated optical systems with balanced gain and loss. We demonstrate that whether the original system is $\mathcal{PT}$-symmetric or not, we can manipulate the property of the $\mathcal{PT}$ symmetry by applying a periodic modulation in such a way that the effective system derived by the high-frequency Floquet method is $\mathcal{PT}$ symmetric. If the original system is non-$\mathcal{PT}$ symmetric, the $\mathcal{PT}$ symmetry in the effective system will lead to quasi-stationary propagation that can be associated  with the \emph{pseudo $\mathcal{PT}$ symmetry}. Our results provide a promising approach for manipulating the $\mathcal{PT}$ symmetry of realistic systems.

\end{abstract}

\maketitle

Parity-time ($\mathcal{PT}$) symmetry, the invariance under parity-time reflection, is an important concept in physics recently developed in application to optical systems. The parity reflection operator $\left(\hat{P}: \hat{x} \rightarrow -\hat{x}, \hat{p} \rightarrow -\hat{p}\right)$ and the time reversal operator $\left(\hat{T}: \hat{x} \rightarrow \hat{x}, \hat{p} \rightarrow -\hat{p}, i \rightarrow -i, t \rightarrow -t\right)$ are defined by their action on the position operator $\hat{x}$, the momentum operator $\hat{p}$ and the time $t$. In quantum mechanics, the requirement of Hermitian Hamiltonians guarantees the existence of real eigenvalues and probability conservation. However, as their Hamiltonian and $\mathcal{PT}$ operators share common eigenfunctions, a wide class of non-Hermitian $\mathcal{PT}$-symmetric Hamiltonians can still possess entirely real eigenvalue spectra~\cite{Bender1,Bender2,Bender3,Bender4}. Although the extension of quantum mechanics based upon non-Hermitian $\mathcal{PT}$-symmetric operators is still a subject to debates, optical systems with complex refractive indices~\cite{Klaiman,El-Ganainy,Sukhorukov,Ramezani, Abdullaev,Makris,Longhi1,Musslimani} are widely used to test the $\mathcal{PT}$ symmetry in non-Hermitian systems, because of the equivalence between the Schr\"{o}dinger equation and the optical wave equation~\cite{Longhi}. In the last few years, the $\mathcal{PT}$ symmetry has been observed in several optical systems, such as optical couplers~\cite{Guo,Ruter}, microwave billiard~\cite{Bittner}, and large-scale temporal lattices~\cite{Regensburger}.

Similar to the electron transport in periodic crystalline potentials and the quantum tunneling in periodically driven systems~\cite{Grifoni}, the light propagation in waveguides can be effectively controlled by periodic modulations~\cite{Valle,Luo,Szameit,Kartashov,Garanovich}. In an optical system, periodic modulation is associated with a periodic refractive index. Mathematically, an optical system with periodic complex refractive index is equivalent to a time-periodic non-Hermitian quantum system. Given the resonant frequency $\omega_0$ for the system without modulation, the modulation frequency $\omega$ and the modulation amplitude $A$, if $\omega_0 \ll \textrm{max}[\omega, \sqrt{|A|\omega}]$, the modulated system can be mapped into an effectively unmodulated one with rescaled parameters~\cite{Grifoni,Shirley}. Like the case of no modulation, the $\mathcal{PT}$ symmetry may appear in the effective system if the periodically modulated system may be described by a $\mathcal{PT}$-symmetric Hamiltonian~\cite{Moiseyev}. Naturally, an important question arises: \emph{Can the $\mathcal{PT}$ symmetry appears in an effective system even if the periodically modulated system is non-$\mathcal{PT}$-symmetric?} In other words, can we employ periodic modulations to manipulate the $\mathcal{PT}$ symmetry?

In this Letter, we study the light propagation in a periodically modulated optical coupler with balanced gain and loss and apply a bi-harmonic modulation along the propagation direction. The Hamiltonian for the modulated system is non-$\mathcal{PT}$-symmetric if the relative phase between the two applied harmonics is not 0 or $\pi$. Applying the high-frequency Floquet approach, the modulated system is effectively described by an effective averaged system, whose $\mathcal{PT}$ symmetry can be manipulated by tuning the modulation amplitude/frequency. More importantly, the $\mathcal{PT}$ symmetry can appear in the effective system corresponding to a non-$\mathcal{PT}$-symmetric and non-Hermitian Hamiltonian. Different from the $\mathcal{PT}$ symmetry from a $\mathcal{PT}$-symmetric Hamiltonian, which leads to stationary light propagation of bounded intensity oscillation, the $\mathcal{PT}$ symmetry from a non-$\mathcal{PT}$-symmetric Hamiltonian will lead to quasi-stationary light propagation of unbounded intensity oscillation. Therefore, we term the induced symmetry associated with modulated systems as the \emph{pseudo $\mathcal{PT}$ symmetry}.

In optics, the electric field $E(x,z)$ of light obeys the wave equation,
\begin{eqnarray}
i\frac{\partial E(x,z)}{\partial z} =-\frac{1}{2k}\frac{\partial^2E(x,z)}{\partial x^2}+V(x,z)E(x,z),
\label{eq:schopt}
\end{eqnarray}
where, $k=k_0 n_0$, $k_0=2\pi/\lambda$ and $V(x,z)=k_0[n_0-n(x)]$ with the substrate index $n_0$, the free-space wavelength $\lambda$ and the complex refractive index distribution $n(x)=n_0+n_R(x,z)+in_I(x)$, where $n_R$ and $n_I$ are real and imaginary parts of $n(x)$. Therefore, the effective potential reads as $V(x,z)=V_R(x,z)+iV_I(x)=-k_0[n_R(x,z)+in_I(x)]$.  With the experimental techniques developed in recent years~\cite{Makris,El-Ganainy,Musslimani,Guo,Ruter}, one can make $V_I(-x)=-V_I(x)$ and  $V_R(x,z)=V_0(x)+V_1(x,z)$ with the unmodulated part $V_0(x)$ being a symmetric double-well function and the modulation $V_1(x,z)=V'(x)F(z)$ described by an anti-symmetric function $V'(-x)=-V'(x)$ and a bi-harmonic function $F(z)$, see Fig.~1.

\begin{figure}[h]
\includegraphics[width=0.8\columnwidth]{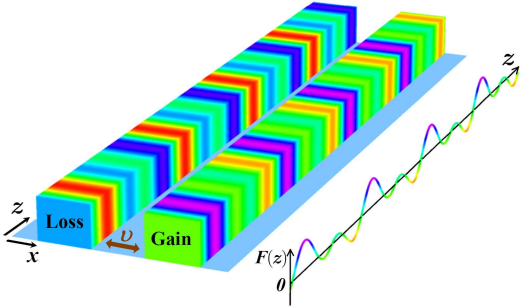}
\caption{Schematic diagram of a modulated two-channel optical coupler with balanced gain and loss. The periodic change of color along $z$-axis denotes the periodic modulation $F(z)$.}
\label{fig:SchematicDiagram}
\end{figure}

Using the coupled-mode theory, the electric field for a two-channel coupler can be expressed as a two-mode ansatz with the localized waves $\left\{\psi_{1}(x),\psi_{2}(x)\right\}$ and the complex amplitudes $\left\{c_{1}(z), c_2(z)\right\}$~\cite{Supplementary}. Thus we have
\begin{eqnarray}
i\frac{d}{dz}\left(\matrix{c_1 \cr c_2} \right) = \left(\matrix{+\frac{i\gamma}{2}+\frac{S(z)}{2} & v \cr v & -\frac{i\gamma}{2}-\frac{S(z)}{2} } \right )\left(\matrix{c_1 \cr c_2} \right),\label{eq1}
\end{eqnarray}
with the inter-channel coupling strength $v$, the gain/loss strength $\gamma$ and the bi-harmonic modulation
\begin{eqnarray}
S(z)=-A[\sin(\omega z)+f\sin(2\omega z+\phi)].
\end{eqnarray}
Here, $\phi\in [0,2\pi)$ denotes the relative phase between the two harmonics, $\omega$ is the modulation frequency, $A$ is the modulation amplitude and $f$ is a dimensionless coefficient. Since the system is invariant under the transformation $c_2\rightarrow-c_2$ and $v\rightarrow-v$, below we will only consider the case of $v>0$. Defining the parity operator as $\hat{P}$, which interchanges the two channels labeled by 1 and 2, and the time operator as $\hat{T}$: $i\rightarrow -i, z\rightarrow -z$, which reverses the propagation direction, the Hamiltonian $\hat{H}$ for the system (2) is $\mathcal{PT}$ symmetric if $\hat{P}\hat{T}\hat{H}=\hat{H}\hat{P}\hat{T}$. If $\phi=0$ or $\pi$, $S(-z)=-S(z)$, $\hat{H}$ is $\mathcal{PT}$-symmetric. Otherwise, if $\phi \neq 0$ and $\pi$, $S(z_0-z)\neq -S(z_0+z)$ for arbitrary constant $z_0$, $\hat{H}$ becomes non-$\mathcal{PT}$-symmetric.

Under the condition of $v \ll \textrm{max}[\omega, \sqrt{|A|\omega}]$, one can implement the high-frequency Floquet analysis. Introducing the transformation
\begin{eqnarray}
c_1=c_1'\exp\Big\{-i\Big[\frac{A}{2\omega} \cos(\omega z)+\frac{Af}{4\omega}\cos(2\omega z+\phi)\Big]\Big\}, \label{trans1}\\
c_2=c_2'\exp\Big\{+i\Big[\frac{A}{2\omega} \cos(\omega z)+\frac{Af}{4\omega}\cos(2\omega z+\phi)\Big]\Big\}\label{trans2},
\end{eqnarray}
and averaging the high frequency terms, one can obtain an effectively unmodulated system
\begin{eqnarray}
i\frac{d}{dz}\left(\matrix{c_1' \cr c_2'} \right)= \left(\matrix{+\frac{i\gamma}{2} & J \cr
J^* & -\frac{i\gamma}{2}} \right)
 \left(\matrix{c_1' \cr c_2'} \right),
\end{eqnarray}
with the rescaled coupling strength
\begin{eqnarray}
J=v\sum_{m=-\infty}^{\infty}(i)^{-m} J_{-2m}\left(\frac{A}{\omega}\right) J_{m}\left(\frac{Af}{2\omega}\right) \exp(im\phi).\label{J}
\end{eqnarray}
The modulus of $J$ depends on the values of $A/\omega$ and $\phi$. If $A/\omega$ is relatively small, the modulus $|J|$ is almost independent on the relative phase $\phi$. When $A/\omega$ increases, the modulus $|J|$ becomes sensitively dependent on $\phi$. In particular, the modulus $|J|$ equals to zero at some specific values of $A/\omega$ (such as $A/\omega \simeq 2.4$ and $5.52$) and $\phi=\pi/2$ or $3\pi/2$. In Fig.~2 (a), choosing $f=1/4$, we show the contour plot of $|J|$ as a function of $A/\omega$ and $\phi$.

By diagonalizing the Hamiltonian for the effective system~(6), the two eigenvalues are given as
\begin{eqnarray}
\varepsilon=\pm|J| \sqrt{1- \left[\gamma/(2J)\right]^2}. \label{eq:eigenvalue}
\end{eqnarray}
Obviously, dependent on the values of $\frac{\gamma}{2|J|}$, the two eigenvalues can be real or complex. The two eigenvalues are real if $\gamma<2|J|$ and they become complex if $\gamma>2|J|$. Therefore, $\gamma_{critical}=2|J|$ is the critical point for the phase transition between real and complex spectra in the effective system, which corresponds to the original system~(2) under high-frequency modulations. The spontaneous $\mathcal{PT}$-symmetry-breaking transition takes place in the effective model~(6) when the imaginary part of $\varepsilon$ changes from zero to nonzero. Surprisingly, unlike our conventional understanding, we find that the quasi-energies can be real even if the modulated system (2) is non-$\mathcal{PT}$-symmetric (i.e. $\phi\ne 0$ and $\pi$).

The parametric dependence of $\left|\textrm{Im}(\varepsilon)\right|$ is shown in Fig.~2 (b-e). In Fig.~2 (b-c), we show $\left|\textrm{Im}(\varepsilon)\right|$ as a function of $\gamma$ and $\phi$ for $f=1/4$. For small $A/\omega$, such as $A/\omega=1$ in Fig.~2 (b), $\left|\textrm{Im}(\varepsilon)\right|$ is almost independent on $\phi$ and the transition from a completely real quasi-energy spectrum ($\left|\textrm{Im}(\varepsilon)\right|=0$) to a complex spectrum ($\left|\textrm{Im}(\varepsilon)\right|\ne 0$) take places when $\gamma$ increases. Near a minimum of $|J|$, such as $A/\omega = 2.4$, $\left|\textrm{Im}(\varepsilon)\right|$ strongly depends on $\phi$, see Fig.~2 (b). In Fig.~2 (d-e), we show $\left|\textrm{Im}(\varepsilon)\right|$ as a function of $\gamma$ and $A/\omega$ for (d) $\phi=0$ and (e) $\phi=\pi/2$. Near the minima of $|J|$, such as $A/\omega\simeq 2.4, 5.52, ...$, $\left|\textrm{Im}(\varepsilon)\right|$ shows significant difference between the two cases of $\phi=0$ and $\phi=\pi/2$. In particular, at the minimum points, $|J|$ vanished to zero for $\phi=\pi/2$ and the corresponding critical value $\gamma_{critical}=2|J|$ is reduced to zero. Similar to a non-Hermitian system with no modulations, the spontaneous $\mathcal{PT}$-symmetry-breaking transition ($\left|\textrm{Im}(\varepsilon)\right|=0 \Rightarrow \left|\textrm{Im}(\varepsilon)\right|\ne 0$) can be observed by tuning the gain/loss strength $\gamma$. More interestingly, for our modulated system~(2) of fixed $\gamma$, it is possible to observe the spontaneous $\mathcal{PT}$-symmetry-breaking transition by tuning $\phi$ and $A/\omega$, see Fig.~2 (c-e).

\begin{figure}[htp]
\center
\includegraphics[width=1.0\columnwidth]{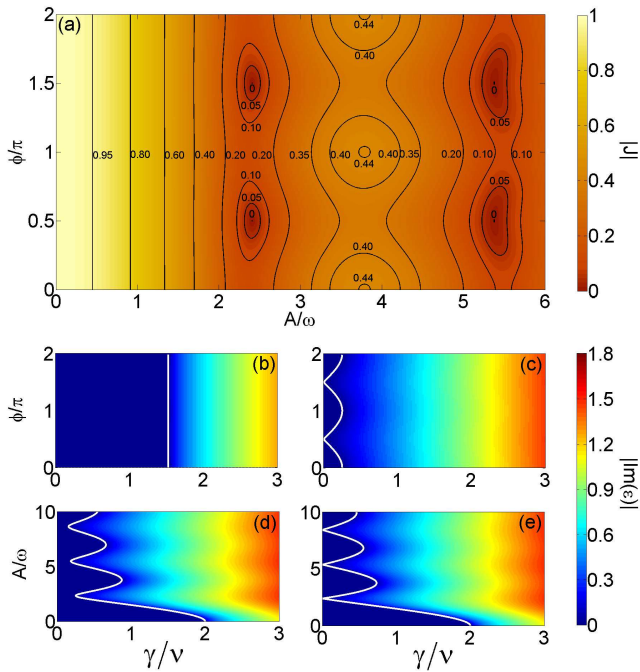}
\caption{(color online) The parametric dependence of the effective coupling $\left|J\right|$ and the imaginary parts of quasi-energies $\left|\textrm{Im}(\varepsilon)\right|$. Top row [(a)]: $\left|J\right|$ versus $A/\omega$ and $\phi$ for $f=1/4$. Middle row [(b) and (c)]: $\left|\textrm{Im}(\varepsilon)\right|$ versus $\phi/\pi$ and $\gamma$ for (b) $A/\omega=1$ and (c) $A/\omega=2.4$. Bottom row [(d) and (e)]: $\left|\textrm{Im}(\varepsilon)\right|$ versus $A/\omega$ and $\gamma$ for (d) $\phi=0$ and (e) $\phi=\pi/2$. The other parameters for $\left|\textrm{Im}(\varepsilon)\right|$ are chosen as $v=1$, $\omega=10$ and $f=1/4$. The white curves are the boundary $\left(\gamma_{critical}=2|J|\right)$ between $\left|\textrm{Im}(\varepsilon)\right|=0$ and $\left|\textrm{Im}(\varepsilon)\right|\ne 0$.} \label{fig2}
\end{figure}

Based upon the high-frequency Floquet analysis, it seems that, whether the modulated system~(2) obeys a $\mathcal{PT}$-symmetric Hamiltonian or not, completely real quasi-energy spectrum always appear if $\gamma<2|J|$. This is obviously inconsistent with the previous theory~\cite{Bender1,Bender2,Bender3,Bender4} which tells us that only $\mathcal{PT}$-symmetric Hamiltonian systems can support completely real spectra. So, what really happens in the modulated non-Hermitian and non-$\mathcal{PT}$-symmetric Hamiltonian system?

In general, according to the Floquet theorem, one can use numerical method to calculate the Floquet states and their quasi-energies for arbitrary modulation frequency and amplitude. Similar to the Bloch states, the Floquet states of the modulated system~(\ref{eq1}) satisfy $\{c_1(z),c_2(z)\}=e^{-i\varepsilon z} \{\tilde{c}_1(z),\tilde{c}_2(z)\}$. Here, the propagation constant $\varepsilon$ is called as the quasi-energy, and the complex amplitudes $\tilde{c}_1(z)$ and $\tilde{c}_2(z)$ are periodic with the modulation period $T=2\pi/\omega$.

\begin{figure}[htp]
\center
\includegraphics[width=1.0\columnwidth]{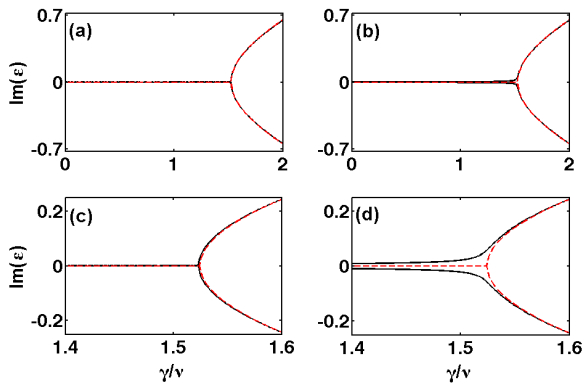}
\caption{(color online) Comparison between numerical and analytical results of $\textrm{Im}(\varepsilon)$, the imaginary part of quasi-energy. Upper row: $\textrm{Im}(\varepsilon)$ versus $\gamma$ for (a) $\phi=0$ and (b) $\phi=\pi/2$. Solid lines are for numerical results obtained from the original model~(\ref{eq1}) and red dashed lines are analytical results given by the formula~(8) for the effective model~(6). Lower row: the enlarged regions of (a) and (b) near the bifurcation point given by the analytical formula~(8). The other parameters are $v=1$, $A=10$, $f=1/4$ and $\omega=10$.} \label{fig3}
\end{figure}

To show the validity of the high-frequency Floquet analysis, we compare the numerical quasi-energies obtained from the original model~(\ref{eq1}) and the analytical formula~(8) obtained from the effective model~(6). In the high-frequency regime, $v \ll \textrm{max}[\omega, \sqrt{|A|\omega}]$, the analytical and numerical values for the quasi-energies $\varepsilon$ are in good agreement and only show tiny difference dependent upon $\phi$. As two examples, we show $\textrm{Im}(\varepsilon)$ (the imaginary part of quasi-energy) versus $\gamma$ for $\phi=0$ and $\phi=\pi/2$ in Fig.~\ref{fig3} (a) and (b), respectively. It clearly shows that the analytical results (red lines) agree well with the numerical results (black lines). Below the critical point ($\gamma < \gamma_{critical}=2|J|$), for $\mathcal{PT}$-symmetric Hamiltonian systems ($\phi=0$ or $\pi$), the numerical results confirm the entirely real quasi-energy spectrum, see Fig.~3~(c). However, for non-$\mathcal{PT}$-symmetric Hamiltonian systems ($\phi \ne 0$ and $\pi$), the numerical quasi-energies $\varepsilon$ still have small non-zero imaginary parts even if $\gamma < \gamma_{critical}=2|J|$, see Fig.~3~(d). This means that, if the original system~(2) obeys a non-$\mathcal{PT}$-symmetric Hamiltonian, the entirely real quasi-energy spectrum for the effective model~(6) does not correspond to perfectly entirely real quasi-energy spectrum for the original system~(2). Therefore, such a $\mathcal{PT}$ symmetry in the effective model~(6) corresponds to a kind of \emph{pseudo $\mathcal{PT}$ symmetry} in the original model~(2). The appearance of \emph{pseudo $\mathcal{PT}$ symmetry} indicates that the deviation of the high-frequency Floquet analysis depends on both the modulation frequency $\omega$ and the Hamiltonian symmetry. Nevertheless, this deviation tends to zero when $\omega \rightarrow \infty$~\cite{Supplementary}.

Through numerical integration, we analyze the light propagations in the continuous system~(1) and the coupled-mode system~(2). The light propagation sensitively depends upon the quasi-energies. Stationary light propagations of bounded intensity oscillations appear if all quasi-energies are real. Non-stationary light propagations of unbounded intensity oscillations appear if at least one of quasi-energies is complex, in which quasi-stationary light propagations of slowly varying time-averaged intensities appear if the two quasi-energies for the effective system~(6) are real. In Fig.~4, for $v=1$, $A=10$, $f=1/4$, $\omega=10$ and $\gamma=0.1$ (which is below the critical value $\gamma_{critical}$), we show the intensity evolution of the coupled-mode system from $c_1(0)=1$ and $c_2(0)=0$. In which, the two intensities $I_{j}(z)=\left|c_{j}(z)\right|^2$, the total intensity $I_{t}(z)=I_1(z)+I_2(z)$ and the time-averaged total intensity $I^{av}_{t}(z)=\frac{1}{T_{s}}\int_{z}^{z+T_{s}} I_{t}(\tilde{z}) d\tilde{z}$ with $T_{s}=2\pi/\left|\textrm{Re}(\varepsilon_{2}) -\textrm{Re}(\varepsilon_{1})\right|$ and $\textrm{Re}(\varepsilon_{j})$ being the real part of $\varepsilon_{j}$. In short-distance propagations, $I_{1,2}(z)$ and $I_{t}(z)$ oscillate periodically and it is hard to see the difference between the cases of $\phi=0$ and $\pi/2$, see Fig. 4 (c-d). However, significant difference appears in long-distance propagations. For a $\mathcal{PT}$-symmetric Hamiltonian system of $\phi=0$, $I^{av}_{t}(z)$ keeps unchanged, see Fig. 4 (a). For a non-$\mathcal{PT}$-symmetric Hamiltonian system of $\phi=\pi/2$, $I^{av}_{t}(z)$ slowly increases, see Fig. 4 (b). The quasi-stationary light propagations of slowly varying $I^{av}_{t}(z)$ is a direct signature of the \emph{pseudo $\mathcal{PT}$ symmetry}. Moreover, our numerical simulations of the continuous wave equation~(1) perfectly confirm the \emph{pseudo $\mathcal{PT}$ symmetry} predicted by the corresponding coupled-mode system~\cite{Supplementary}.

\begin{figure}[htp]
\center
\includegraphics[width=1.0\columnwidth]{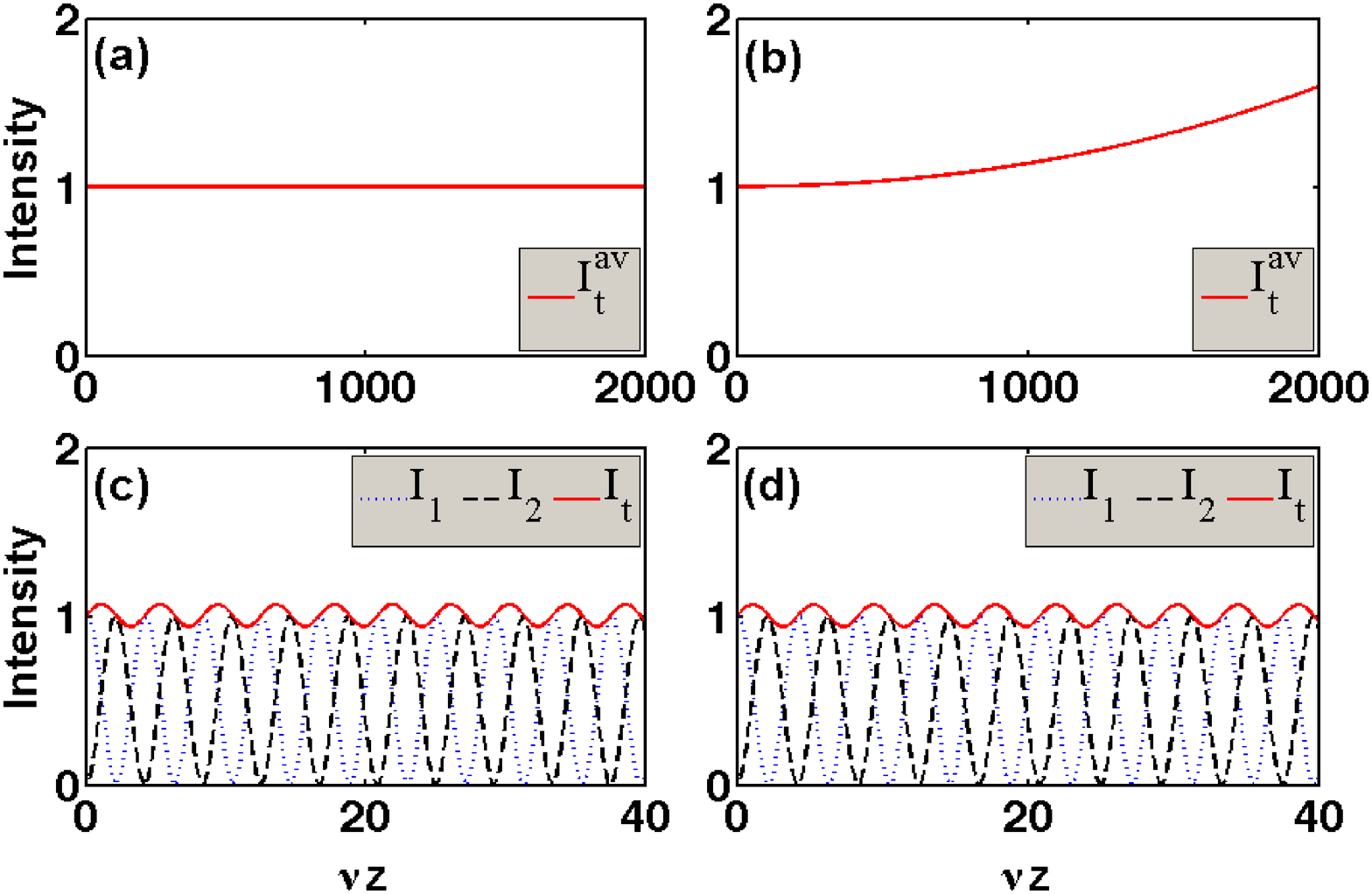}
\caption{(color online) Intensity evolution from the initial state of $c_1(0)=1$ and $c_2(0)=0$. Upper row : long-distance time-averaged intensity evolution for (a) $\phi=0$ and (b) $\phi=\pi/2$. Lower row: short-distance intensity evolution for (c) $\phi=0$ and (d) $\phi=\pi/2$. The other parameters are chosen as $v=1$, $A=10$, $f=1/4$, $\omega=10$ and $\gamma=0.1$.} \label{fig4}
\end{figure}

Now, we discuss the experimental possibility to observe our theoretical predictions. Recently, several $\mathcal{PT}$-symmetric optical systems have been realized experimentally~\cite{Makris,El-Ganainy,Musslimani,Guo,Ruter}. Complex refractive index of gain/loss effects can be obtained from quantum-well lasers or photorefractive structures through two-wave mixing~\cite{Yariv-Yeh}. Periodic modulations can be introduced by out-of-phase harmonic modulations of the real refractive index~\cite{Szameit,Longhi,Garanovich} or periodic curvatures along the propagation direction~\cite{Valle,Zeuner,Longhi,Garanovich}. For a short optical coupler under periodic modulations, spontaneous $\mathcal{PT}$-symmetry-breaking transitions can be observed, whether the system Hamiltonian is $\mathcal{PT}$-symmetric or not. In such a system, the light propagation will be periodic and stable if $\gamma<2|J|$, and an instability will be observed if $\gamma>2|J|$. The critical point $\gamma_{critical}=2|J|$ can be adjusted by controlling the modulation parameters $A/\omega$, $f$ and $\phi$ in addition to controlling $\gamma$. However, for a long optical coupler under periodic modulations, the light propagation below the critical point ($\gamma<2|J|$) depends on the Hamiltonian symmetry. If the Hamiltonian is $\mathcal{PT}$-symmetric, the light propagation is periodic and stable, in which the time-averaged total intensity keeps unchanged. Otherwise, if the Hamiltonian is non-$\mathcal{PT}$-symmetric, the light propagation is quasi-stationary, in which the time-averaged total intensity slowly grows.

In summary, we have studied the non-Hermitian Hamiltonian systems under periodic modulations and introduce the concept of the \emph{pseudo $\mathcal{PT}$ symmetry}. If the modulated system obeys a $\mathcal{PT}$ symmetric Hamiltonian, there exists a truly spontaneous $\mathcal{PT}$-symmetry-breaking phase transition from a real quasi-energy spectrum to a complex one. If the modulated system obeys a non-$\mathcal{PT}$-symmetric Hamiltonian, although there exists a spontaneous $\mathcal{PT}$-symmetry-breaking phase transition in the effective system derived from the high-frequency Floquet analysis, there is no truly spontaneous $\mathcal{PT}$-symmetry-breaking phase transition in the original system. Corresponding to the real spectrum for the effective system, the original system has a quasi-real spectrum of small imaginary parts, which leads to a quasi-stationary light propagation of slowly varying time-averaged total intensity. This is the \emph{pseudo $\mathcal{PT}$ symmetry} in the non-Hermitian system described by a non-$\mathcal{PT}$-symmetric Hamiltonian.

In addition to the discovery of the \emph{pseudo $\mathcal{PT}$ symmetry}, we believe that our work brings three key advances to related fields. Firstly, although the high-frequency Floquet analysis can capture most key features, some important information (such as the so-called \emph{pseudo $\mathcal{PT}$ symmetry}) may be lost. Secondly, periodic modulations provide a new route to the observation of the spontaneous $\mathcal{PT}$-symmetry-breaking transition. Thirdly, as the inter-channel coupling can be effectively switched off by controlling the modulation and the corresponding intensity grows exponentially even for arbitrarily weak gain/loss, this exponential growth offers an efficient way to beam amplification in optical waveguides.

X. Luo and J. Huang have made equal contributions. This work is supported by the NBRPC under Grant No. 2012CB821305, the NNSFC under Grants No. 11075223, 11165009 and 10965001, the Ph.D. Programs Foundation of Ministry of Education of China under Grant No. 20120171110022, and the NCETPC under Grant No. NCET-10-0850. X. Luo is also partially supported by the Natural Science Foundation of Jiangxi Province under Grant No. 2010GQW0033, the Jiangxi Young Scientists Training Plan under Grant No. 20112BCB23024 and the Key Subject of Atomic and Molecular Physics in Jiangxi Province.

\section{Supplementary Material}

\subsection{Coupled-mode theory}

In our system, the electric field $E(x,z)$ of light obeys
\begin{equation}
i\frac{\partial E(x,z)}{\partial z}=-\frac{1} {2k}\frac{\partial^{2} E(x,z) }{\partial x^2}+V(x,z)E(x,z), \label{waveeq}
\end{equation}
with the refractive index, $V(x,z)=V_R(x,z)+iV_I(x)$. The real part $V_R(x,z)=V_0(x)+V_1(x,z)$ with the symmetric double-well function $V_0(x)$ and the periodic modulation $V_1(x,z)=V'(x)F(z)$. Here, $V'(-x)=-V'(x)$ and the imaginary part $V_I(-x)=-V_I(x)$ are anti-symmetric functions, and $F(z)$ is periodic function.

To compare with the Schr\"{o}dinger equation, we rewrite the continuous wave equation (1) as
\begin{equation}
i\frac{\partial E(x,z)}{\partial z}=\hat{H}E(x,z),
\end{equation}
with
\begin{eqnarray}
\hat{H} &=& \hat{H}_0+V'(x)F(z)+iV_I(x),\nonumber\\
\hat{H}_0 &=& -\frac{1}{2k}\frac{\partial^{2}}{\partial x^2} +V_0(x) = \frac{\hat{p}^2} {2k}+V_0(x). \nonumber
\end{eqnarray}
The parity operator $\hat{P}$: $x\rightarrow -x$ and $\hat{p}\rightarrow -\hat{p}$, has the effect of reversing the transverse direction. The time operator $\hat{T}$: $x\rightarrow x$, $\hat{p}\rightarrow -\hat{p}$, $i\rightarrow -i$ and $z\rightarrow -z$, has the effect of reversing the propagation direction. The Hamiltonian $\hat{H}$ is $\mathcal{PT}$-symmetric if $\hat{P}\hat{T}\hat{H}=\hat{H}\hat{P}\hat{T}$.

If the two wells of $V_0(x)$ are sufficiently deep, the electric field can be expressed as the two-mode ansatz
\begin{equation}
E(x,z)=[c_1(z)\psi_1(x)+c_2(z)\psi_2(x)]\exp({-i\epsilon_0z}), \label{appro}
\end{equation}
where
\begin{eqnarray}
\psi_1(x)&=&\frac{1}{\sqrt{2}}\left(\phi_g(x)+\phi_e(x)\right),\nonumber\\ \psi_2(x)&=&\frac{1}{\sqrt{2}}\left(\phi_g(x)-\phi_e(x)\right),\nonumber
\end{eqnarray}
are the two localized waves in the two wells of $V_0(x)$. Here, $\phi_{g}(x)$ and $\phi_{e}(x)$ are the two lowest eigenstates for $\hat{H}_0$, that is,
\begin{eqnarray}
\hat{H}_0\phi_{g}(x)&=&\epsilon_{g}\phi_{g}(x),\nonumber\\
\hat{H}_0\phi_{e}(x)&=&\epsilon_{e}\phi_{e}(x).\nonumber
\end{eqnarray}
It is easy to find $\psi_{1}$ and $\psi_{2}$ satisfy the orthonormal condition, $\int_{-\infty}^{+\infty}\psi_{i}^*(x)\psi_{j}(x)dx = \delta_{ij}$.

By using the two-mode ansatz~(\ref{appro}), Eq.~(\ref{waveeq}) becomes
\begin{eqnarray}
\epsilon_0(c_1\psi_1+c_2\psi_2) +i\dot{c_1}\psi_1+i\dot{c_2}\psi_2 =\hat{H}(c_1\psi_1+c_2\psi_2), \label{der1}
\end{eqnarray}
with $\epsilon_0=\int_{-\infty}^{+\infty} \psi_{1,2}^*(x)\hat{H}_0\psi_{1,2}(x)dx =\frac{1}{2} \left(\epsilon_g+\epsilon_e\right)$.
Multiplying $\psi_{1,2}^*(x)$ on both sides of Eq.~(\ref{der1}) and integrating both sides with respect to $x$ from $-\infty$ to $+\infty$, we have
\begin{eqnarray}
\epsilon_0c_1+i\dot{c_1}=c_1\int_{-\infty}^{\infty} \psi_1^*\hat{H}_0\psi_1dx+c_2\int_{-\infty}^{+\infty} \psi_1^*\hat{H}_0\psi_2dx \nonumber\\
+F(z)c_1\int_{-\infty}^{+\infty} \psi_1^*V'(x)\psi_1dx+F(z)c_2\int_{-\infty}^{+\infty} \psi_1^*V'(x)\psi_2dx\nonumber \\
+ic_1\int_{-\infty}^{+\infty} \psi_1^*V_I
\psi_1dx+ic_2\int_{-\infty}^{+\infty} \psi_1^*V_I \psi_2dx,\nonumber
\end{eqnarray}
\begin{eqnarray}
\epsilon_0c_2+i\dot{c_2}=c_1\int_{-\infty}^{+\infty} \psi_2^*\hat{H}_0\psi_1dx +c_2\int_{-\infty}^{+\infty}\psi_2^*\hat{H}_0\psi_2dx \nonumber\\
+F(z)c_1\int_{-\infty}^{+\infty} \psi_2^*V'(x)\psi_1dx+F(z)c_2\int_{-\infty}^{+\infty} \psi_2^*V'(x)\psi_2dx\nonumber \\
+ic_1\int_{-\infty}^{+\infty} \psi_2^*V_I
\psi_1dx+ic_2\int_{-\infty}^{+\infty}\psi_2^*V_I \psi_2dx. \nonumber
\end{eqnarray}
Without loss of generality, one can assume the two localized waves $\psi_{1,2} (x)$ are real functions. By using the relations, $\psi_1(-x)=\psi_2(x)$, $\psi_1(-x)\psi_2(-x)=\psi_1(x)\psi_2(x)$, $V'(-x)=-V'(x)$, and $V_I(-x)=-V_I(x)$,
we have
\begin{eqnarray}
\int_{-\infty}^{+\infty}\psi_1^*\hat{H}_0\psi_2dx &=&\int_{-\infty}^{+\infty}\psi_2^*\hat{H}_0\psi_1dx =-\frac{\epsilon_{e}-\epsilon_{g}}{2},\nonumber \\
\int_{-\infty}^{+\infty} \psi_2^*V'(x)\psi_1dx &=&\int_{-\infty}^{+\infty} \psi_1^*V'(x)\psi_2dx=0,\nonumber \\
\int_{-\infty}^{+\infty} \psi_1^*V'(x)\psi_1dx &=&-\int_{-\infty}^{+\infty} \psi_2^*V'(x)\psi_2dx,\nonumber\\
\int_{-\infty}^{+\infty} \psi_1^*V_I(x) \psi_1dx &=&-\int_{-\infty}^{+\infty} \psi_1^*V_I(x) \psi_1dx.\nonumber
\end{eqnarray}
After some mathematical iteration, we obtain the coupled-mode equations,
\begin{eqnarray}
i\frac{d}{dz} \left(\matrix{c_1 \cr c_2}\right) = \left(\matrix{+\frac{i\gamma}{2}+\frac{S(z)}{2} & v \cr v & -\frac{i\gamma}{2}-\frac{S(z)}{2} } \right )\left(\matrix{c_1 \cr c_2} \right),\label{tmaeq}
\end{eqnarray}
for the two field amplitudes $\left\{c_{1}(z), c_2(z)\right\}$. Here, the parameters are given as
\begin{eqnarray}
v&=&\int_{-\infty}^{\infty} \psi_1^*(x)H_0\psi_2(x)dx =-\frac{\epsilon_{e}-\epsilon_{g}}{2},\label{coupling}\\
\gamma&=&2\int_{-\infty}^{\infty} \psi_1^*(x)V_I(x) \psi_1(x)dx,\label{loss}\\
S(z)&=&2F(z)\int_{-\infty}^{\infty} \psi_1^*(x)V'(x)
\psi_1(x)dx.\label{driving}
\end{eqnarray}
Obviously, the coupled-mode equations~(\ref{tmaeq}) are invariant under the transformation $c_2\rightarrow-c_2$ and $v\rightarrow-v$.

\subsection{Numerical simulation of the continuous wave equation}

To confirm the validity of the coupled-mode theory, we directly simulate the continuous wave equation (\ref{waveeq}) and compare with the results predicted by the corresponding coupled-mode equations (\ref{tmaeq}). In our numerical simulations, we choose
\begin{eqnarray}
V_0(x)&=&-p \left(\exp\left[-(\frac{x+\frac{w_s}{2}}{w_x})^6\right] +\exp\left[-(\frac{x-\frac{w_s}{2}}{w_x})^6\right]\right), \nonumber\\
V'(x)&=&-p\mu \left(\exp\left[-(\frac{x+\frac{w_s}{2}}{w_x})^6\right] -\exp\left[-(\frac{x-\frac{w_s}{2}}{w_x})^6\right]\right), \nonumber\\
V_I(x)&=&-p\alpha \left(\exp\left[-(\frac{x+\frac{w_s}{2}}{w_x})^6\right] -\exp\left[-(\frac{x-\frac{w_s}{2}}{w_x})^6\right]\right), \nonumber\\
F(z)&=& \sin(\omega z)+f\sin(2\omega z+\phi). \nonumber
\end{eqnarray}
Here $w_s$ is the distance between two waveguides, $w_x$ is the waveguide width, $(p, \mu, \alpha)$ are three parameters describing the refractive index, and $F(z)$ denotes the bi-harmonic modulation of the refractive index. From Eq.~(\ref{driving}), it is easy to obtain
\begin{eqnarray}
S(z)=-A[\sin(\omega z)+f\sin(2\omega z+\phi)],
\end{eqnarray}
with
\begin{equation}
A=-2 \int_{-\infty}^{+\infty} \psi_1^*(x)V'(x)
\psi_1(x)dx.
\end{equation}
To estimate the parameters $v$, $\gamma$ and $A$ for the corresponding coupled-mode system~(\ref{tmaeq}), we firstly find the two lowest states $\phi_{g,e}(x)$ via imaginary-time evolution and then construct the two localized waves $\psi_{1,2}(x)$.

Below, we discuss the numerical results obtained by directly integrating the the continuous wave equation (1). For simplicity and without loss of generality, we choose the input state as $E(x,0)=\psi_{1}(x)$ and set the parameter $k=1$. Fixing $w_s=3.2$, $w_x=0.3$, $p=3$, $\mu=0.07$, $\alpha=0.014$, $f=0.25$ and $\omega=0.2168$, we simulate the continuous wave equation (1) for different values of $\phi$. As in Refs.~[Phys. Rev. Lett. 102, 153901 (2009); Opt. Lett. 34, 2700 (2009)], $w_s$ and $w_x$ are in units of $10\mu m$, and $p=3$ corresponds to a refractive index of $\sim 4 \times 10^{-4}$.

\begin{figure}[htb]
\includegraphics[width=1.0\columnwidth]{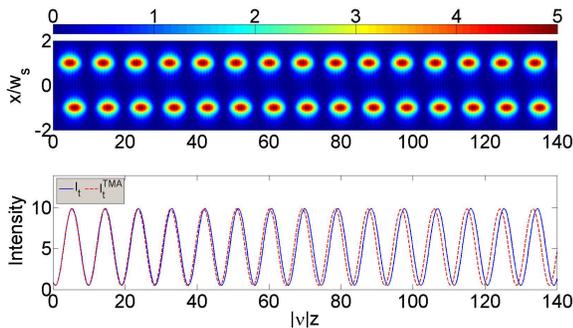}
\caption{Stationary periodic oscillations in the modulated two-channel coupler obeying a $\mathcal{PT}$-symmetric Hamiltonian. Top row: the electric field density $|E(x,z)|^2$ given by the continuous wave equation (1). Bottom row: the total intensity $I_t(z)$ (solid curve) corresponding to the top row and the total intensity $I_t^{TMA}(z)$ (dashed curve) given by the corresponding two-mode system (\ref{tmaeq}).}
\label{fig:PTS-CWE}
\end{figure}

In Fig. 1, we show the numerical results for the continuous wave equation (1) of $\phi=0$, which corresponds to a $\mathcal{PT}$-symmetric Hamiltonian. The corresponding two-mode system (\ref{tmaeq}) has $v=-0.0280$, $\gamma=-0.0400$, $A=0.1999$, $\left|J/v\right|=0.7959$ and $\left|\gamma/J\right|=1.7958$. The electric field density $|E(x,z)|^2$ (top row) and the total intensity $I_t(z)=\int_{-\infty}^{\infty}|E(x,z)|^2dx$ (solid curve in bottom row) show stationary periodic oscillations along the propagation direction. By inputting the estimated parameters into the two-mode system (\ref{tmaeq}), we find that its dynamics is perfectly consistent with the ones predicted by the continuous wave equation (1). In the bottom row, we show the two total intensities together. The total intensity given by the two-mode system (\ref{tmaeq}), $I_t^{TMA}(z)=\left|c_1(z)\right|^2+\left|c_2(z)\right|^2$ (dashed curve in bottom row), is well consistent with the total intensity given by the continuous wave equation (1), $I_t(z)=\int_{-\infty}^{\infty}|E(x,z)|^2dx$ (solid curve in bottom row). The stationary periodic oscillations confirms the existence of $\mathcal{PT}$-symmetry in the region of $\left|\gamma/J\right|<2$.

\begin{figure}[htb]
\includegraphics[width=1.0\columnwidth]{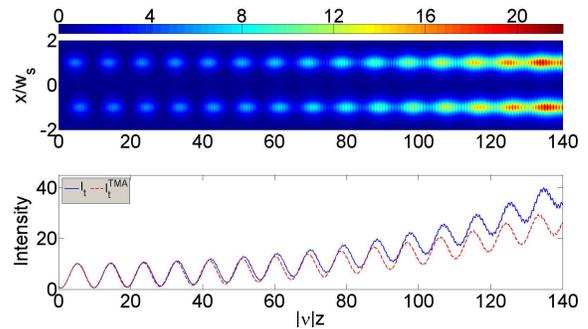}
\caption{Quasi-stationary oscillations in the modulated two-channel coupler obeying a non-$\mathcal{PT}$-symmetric Hamiltonian. Top row: the electric field density $|E(x,z)|^2$ given by the continuous wave equation (1). Bottom row: the total intensity $I_t(z)$ (solid curve) corresponding to the top row and the total intensity $I_t^{TMA}(z)$ (dashed curve) given by the corresponding two-mode system (\ref{tmaeq}).}
\label{fig:PTSB-CWE}
\end{figure}

In Fig. 2, we show the numerical results for the continuous wave equation (1) of $\phi=\pi/2$, which corresponds to a non-$\mathcal{PT}$-symmetric Hamiltonian. The corresponding two-mode system (\ref{tmaeq}) has $v=-0.0280$, $\gamma=-0.0400$, $A=0.1999$, $\left|J/v\right|=0.7958$ and $\left|\gamma/J\right|=1.7961$. In contrast to the case of a $\mathcal{PT}$-symmetric Hamiltonian, although $\left|\gamma/J\right|<2$, the electric field density $|E(x,z)|^2$ (top row) and the total intensity $I_t(z)$ (solid curve in bottom row) show quasi-stationary oscillations with the tendency of growing total intensity. The total intensity $I_t^{TMA}(z)$ (dashed curve in bottom row) given by the corresponding two-mode system (\ref{tmaeq}) is also well consistent with the total intensity $I_t(z)$ (solid curve in bottom row) given by the continuous wave equation (1). The quasi-stationary oscillations in the region of $\left|\gamma/J\right|<2$ are a signature of the \emph{pseudo $\mathcal{PT}$ symmetry}.

Although two total intensities show slowly growing small difference in long-distance propagations, the physical pictures given by the continuous wave equation and the coupled-mode equations have no difference. One main source of this small difference is the small variation of localized states $\psi_{1,2}(x)$ caused by the periodic modulation.

\subsection{Validity of the high-frequency Floquet analysis}

In general, Floquet states may exist in a periodically modulated system for arbitrary modulation amplitude and frequency. Therefore, in principle, the Floquet approach (which aims to find the Floquet states) is valid for arbitrary modulation amplitude and frequency. However, for arbitrary modulation amplitude and frequency, it is very difficult to obtain analytical results and numerical approach has to be used. In the high-frequency regime, $\omega_0 \ll max[\omega, \sqrt{|A|\omega}]$, the high frequency terms can be averaged and the modulated system can be described by an effectively unmodulated one with rescaled parameters [Phys. Rev. 138, B979 (1965); Phys. Rep. 304, 229 (1998)]. Here, $\omega$ and $A$ denote the modulation frequency and amplitude respectively, and $\omega_0$ is the resonant frequency for the system without modulation.

In our Letter, the coupled-mode system under high-frequency periodic modulation is mapped into the effectively unmodulated system with the rescaled coupling strength given by the analytical formula. Without implementing the high-frequency Floquet analysis, it is impossible to obtain the analytical results for the eigenvalues of the original system. However, it is easily to derive the analytical formula for the eigenvalues of the effective unmodulated system given by the high-frequency Floquet analysis.

\begin{figure}[h]
\includegraphics[width=1.0\columnwidth]{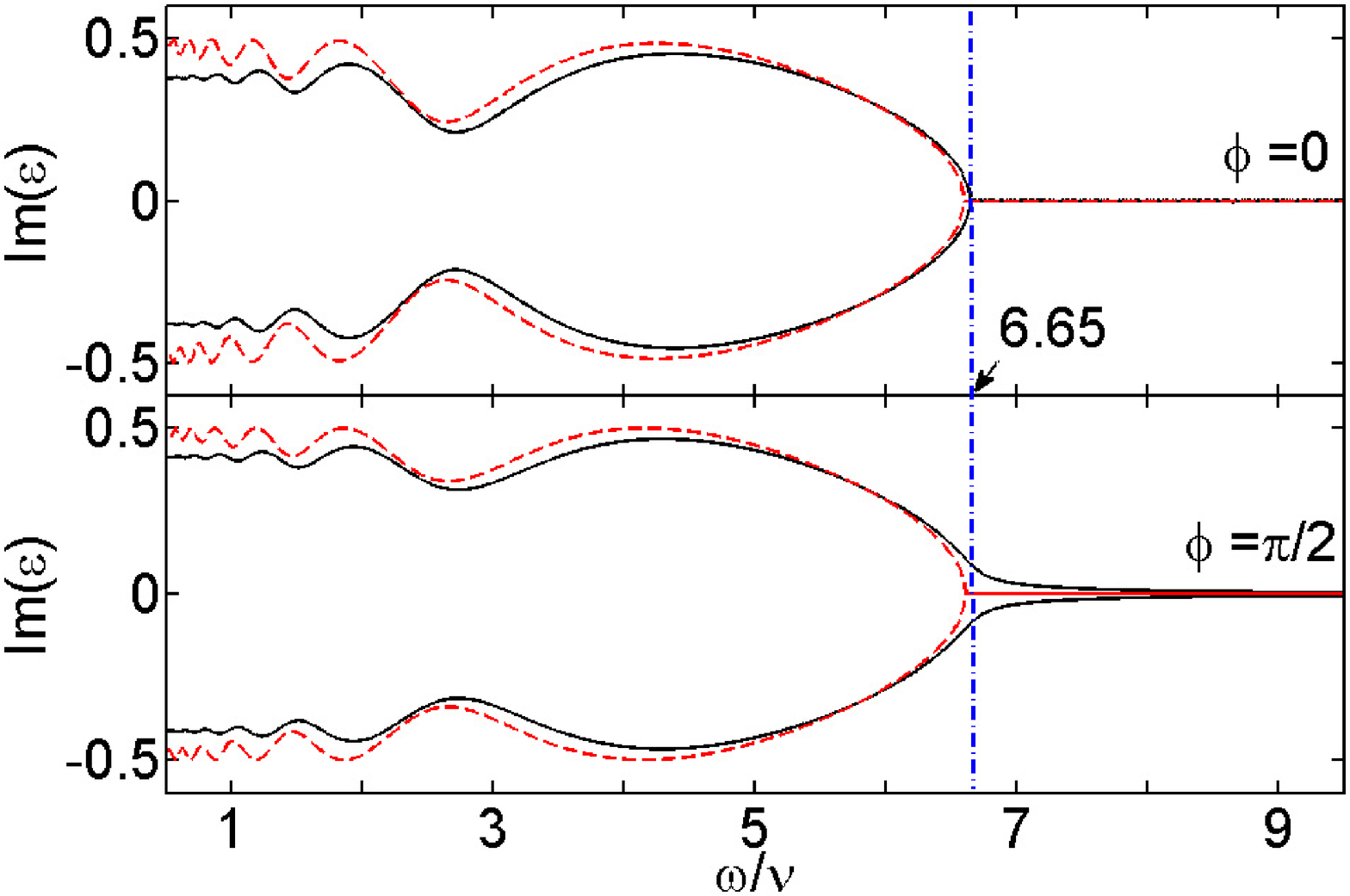}
\caption{The imaginary part of quasi-energy $\textrm{Im}(\epsilon)$ versus the modulation frequency $\omega$. The solid curves are numerical results obtained from the coupled-mode equations~(\ref{tmaeq}) without implementing the high-frequency Floquet analysis. The dashed curves are the analytical results obtained from the high-frequency Floquet analysis [Eq.~(8) in the Letter]. The top and bottom rows correspond to $\phi=0$ and $\pi/2$, respectively. The other parameters are chosen as $\nu=1$, $A=10$, $f=1/4$ and $\gamma=1$.}
\label{ImQuasiE}
\end{figure}

To show the validity of the high-frequency Floquet analysis, we compare the numerical results of the quasi-energies $\varepsilon$ obtained from the original system (solid curves) with the analytical ones obtained from the high-frequency Floquet analysis (dashed curves). In the Letter, we have shown the $\gamma$-dependence of $\textrm{Im}(\varepsilon)$ (the imaginary part of $\varepsilon$) for $\phi=(0,\pi/2)$. Here, we show the $\omega$-dependence of $\textrm{Im}(\varepsilon)$ for $\phi=(0,\pi/2)$ in Fig.~\ref{ImQuasiE}. For $\phi=0$, the numerical and analytical results are almost the same if $|\omega/v| > 6.65$. This indicates the validity of the high-frequency Floquet analysis in the region of $|\omega/v| > 6.65$. However, for $\phi=\pi/2$, although the effectively unmodulated system has $\mathcal{PT}$-symmetry (i.e. $\textrm{Im}(\varepsilon)=0$) in the region of $|\omega/v| > 6.65$, the original system still has a small nonzero $\textrm{Im}(\varepsilon)$ and it tends to zero when $\omega \rightarrow \infty$. The $\mathcal{PT}$-symmetry in the effectively unmodulated system corresponds to a kind of \emph{pseudo $\mathcal{PT}$ symmetry} in the original system. The appearance of \emph{pseudo $\mathcal{PT}$ symmetry} indicates that the deviation of the high-frequency Floquet analysis depends on both the modulation frequency $\omega$ and the Hamiltonian symmetry. In particular, the deviation of the high-frequency Floquet analysis is significant near the critical point between real and complex quasi-energies.

\end{document}